\documentclass[12pt]{iopart}

\usepackage{iopams}  

\expandafter\let\csname equation*\endcsname\relax
\expandafter\let\csname endequation*\endcsname\relax

\usepackage{epsfig, amsmath, amssymb}
\newtheorem{theorem}{Theorem}

\newtheorem{corollary}[theorem]{Corollary}
\usepackage{hyperref}
\usepackage{lineno}

\usepackage{xcolor}
\usepackage{cite}

\begin{document}

\title[A Tensor-Train-Guided Hypernetwork for Robust and Scalable VQC]{TensorHyper-VQC: A Tensor-Train-Guided Hypernetwork for Robust and Scalable Variational Quantum Computing}

\author{Jun Qi$^{1*}$, Chao-Han Huck Yang$^{2}$, Pin-Yu Chen$^{3*}$, Min-Hsiu Hsieh$^{4*}$}

\address{1. School of Electrical and Computer Engineering, Georgia Institute of Technology, Atlanta, GA 30332, USA     \\ 
2. NVIDIA Research, Santa Clara, CA 95051, USA	 \\
3. IBM Thomas J. Watson Research Center, NY, 10598, USA    \\
4. Hon Hai (Foxconn) Quantum Computing Research Center, Taipei, 114, Taiwan   \\
}
\ead{jqi41@gatech.edu, pin-yu.chen@ibm.com, min-hsiu.hsieh@foxconn.com}
\hspace{25mm}\small{* denotes corresponding authors}
\vspace{10pt}


\begin{abstract}
Variational Quantum Computing (VQC) faces fundamental scalability barriers, primarily due to barren plateaus and sensitivity to quantum noise. To address these challenges, we introduce TensorHyper-VQC, a novel tensor-train (TT)-guided hypernetwork framework that significantly improves the robustness and scalability of VQC. Our framework fully delegates the generation of quantum-circuit parameters to a classical TT network, thereby decoupling optimization from quantum hardware. This innovative parameterization mitigates gradient vanishing, enhances noise resilience through structured low-rank representations, and facilitates efficient gradient propagation. Grounded in Neural Tangent Kernel and statistical learning theory, our rigorous theoretical analyses establish strong guarantees on approximation capability, optimization stability, and generalization performance. Extensive empirical results across quantum dot classification, Max-Cut optimization, and molecular quantum simulation tasks demonstrate that TensorHyper-VQC consistently achieves superior performance and robust noise tolerance, including hardware-level validation on a 156-qubit IBM Heron processor. These results position TensorHyper-VQC as a scalable and noise-resilient framework for advancing practical quantum machine learning on near-term devices.
\end{abstract}

%
%
%
%
%

\section{Introduction}
\label{sec1}

Variational Quantum Computing (VQC)~\cite{cerezo2021variational} has emerged as a prominent paradigm for exploiting the computational capabilities of Noisy Intermediate-Scale Quantum (NISQ) devices~\cite{preskill2018quantum}. VQC integrates quantum circuits and classical optimization routines to address computationally challenging problems across various domains~\cite{biamonte2017quantum, cerezo2022challenges, caro2022generalization, mcclean2016theory, power_data, schuld2019quantum}. Despite its significant promise, the practical realization of VQC's potential is substantially hindered by fundamental obstacles related to its trainability, robustness against quantum noise, and overall scalability as problem sizes increase~\cite{holmes2022connecting}. A primary impediment to scalable VQC is the barren plateau phenomenon~\cite{larocca2025barren, mcclean2018barren, martin2023barren}, where gradients vanish exponentially with increasing qubit count or circuit depth, significantly degrading result fidelity and destabilizing training~\cite{beer2020training, sharma2022trainability, pesah2021absence}. 

\begin{figure}
\centerline{\epsfig{figure=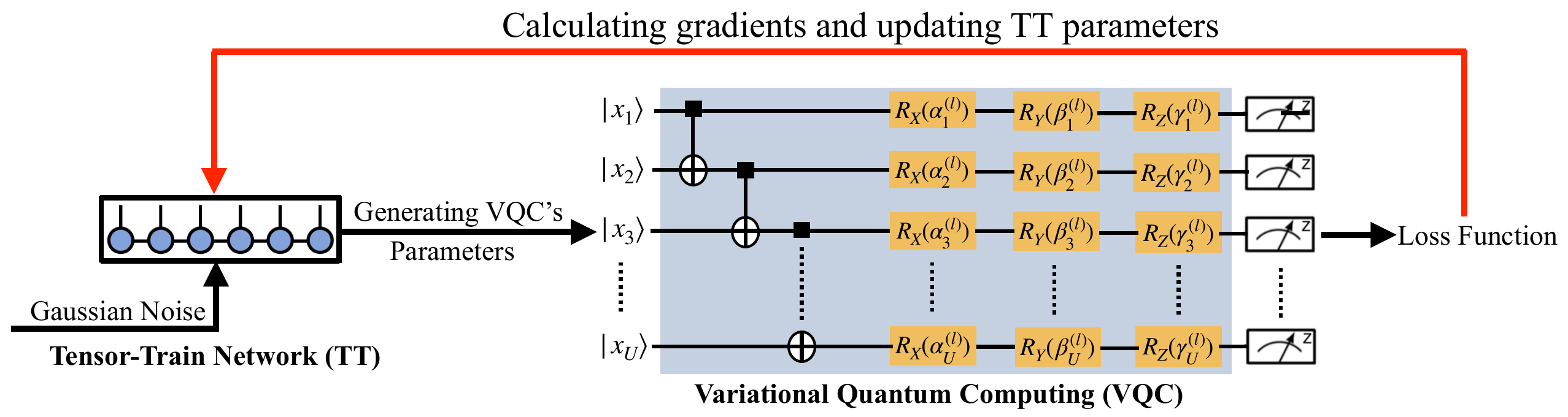, width=130mm}}
\caption{{\it An illustration of the TensorHyper-VQC framework}. A classical TT network generates the parameters $\hat{\textbf{w}} = [\alpha^{(1)}_{1:U}, \beta^{(1)}_{1:U}, \gamma^{(1)}_{1:U}, ..., \alpha^{(L)}_{1:U}, \beta^{(L)}_{1:U}, \gamma^{(L)}_{1:U}]^{\top}$ that define the Pauli rotation gates $R_{X}(\alpha^{(l)}_{u}), R_{Y}(\beta^{(l)}_{u}), R_{Z}(\gamma^{(l)}_{u})$ (where $u\in [U], l\in [L]$) in a VQC with $L$ depths and $U$ qubits. The TT-generated parameters are injected into the fixed quantum circuit, which operates solely in inference mode. The loss is computed from quantum measurements, and gradients are backpropagated only through the classical TT-cores. Gaussian noise is injected during training as the input to the TT network. This architecture decouples optimization from quantum hardware, mitigating barren plateaus and improving noise resilience.}
\label{fig:tt2vqc}
\end{figure}

To overcome these critical limitations, as shown in Fig.~\ref{fig:tt2vqc}, we introduce TensorHyper-VQC, a tensor-train (TT)-guided hypernetwork approach designed to facilitate scalable and robust VQC. The central innovation of TensorHyper-VQC lies in its parameterization strategy. Instead of directly optimizing parameters on the quantum device, a classical TT network~\cite{oseledets2011tensor, yang2017tensor, qi2023exploiting} is trained to generate the parameters for the VQC. Consequently, the entire optimization loop, including gradient computation and parameter updates, is offloaded to the classical TT-cores. In this architecture, the quantum circuit serves solely as a fixed forward-pass evaluator, taking the parameters supplied by the TT network to compute the objective function value. This reparameterization strategy enables a clear separation between classical optimization and quantum inference, ensuring consistent gradient propagation even in the presence of realistic hardware noise.

This hypernetwork-based parameterization fundamentally reshapes the VQC optimization landscape, offering distinct advantages in terms of scalability and robustness. By confining the optimization to the classical TT network, TensorHyper-VQC significantly mitigates the barren-plateau problem, as gradients are backpropagated through the well-conditioned, deterministic structure of the TT network rather than through the potentially deep and complex quantum circuit. The framework's resilience to quantum noise is also markedly enhanced because the trainable parameters reside entirely in the classical domain, insulating gradient updates from the stochasticity of quantum measurements and hardware imperfections. Moreover, the low-rank structure inherent in the TT representation further contributes to noise suppression by effectively averaging measurement noise across the TT-core gradients, a phenomenon theoretically linked to variance reduction. Our Neural Tangent Kernel (NTK)~\cite{bietti2019inductive, jacot2018neural, liu2022representation} analysis supports the improved trainability afforded by this structured parameterization.

Our theoretical investigations confirm the enhanced representation capability, trainability, and generalization properties of TensorHyper-VQC. Experimentally, TensorHyper-VQC demonstrates superior performance and robustness across diverse applications, including quantum machine learning (e.g., quantum dot classification~\cite{kalantre2019machine, czischek2021miniaturizing}), combinatorial optimization (e.g., Max-Cut~\cite{zhou2020quantum, zhu2022adaptive}), and quantum simulation (e.g., LiH Hamiltonians~\cite{o2016scalable, georgescu2014quantum}). In all cases, TensorHyper-VQC outperforms conventional VQC and exhibits strong noise resilience across various noise models. These findings establish TensorHyper-VQC as a powerful and versatile computational framework that advances state-of-the-art variational quantum computing.

\section{Results}
\label{sec2}

\subsection{The TensorHyper-VQC Framework}
Our proposed framework, TensorHyper-VQC, fundamentally re-engineers the VQC training paradigm to address key challenges of scalability and robustness. As illustrated in Figure~\ref{fig:tt2vqc}, the central innovation is to delegate parameter generation to a classical TT network. In this hypernetwork design, the TT network, which is composed of interconnected low-rank tensor cores, serves as the sole trainable component.

Given the VQC depth $L$ and qubit count $U$, the TT network learns to generate the high-dimensional parameter vector as:
\begin{equation}
\hat{\textbf{w}} = [\hat{w}_{1}, \hat{w}_{2}, ..., \hat{w}_{3UL}]^{\top} = [\alpha_{1:U}^{(1)}, \beta_{1:U}^{(1)}, \gamma_{1:U}^{(1)}, ..., \alpha_{1:U}^{(L)}, \beta_{1:U}^{(L)}, \gamma_{1:U}^{(L)}]^{\top}, 
\end{equation}
which defines the unitary rotation gates of the VQC. This reparameterization shifts the entire optimization process into the classical domain, while the quantum circuit is used only for forward evaluation. The overall operational flow is summarized as two decoupled components: a parameter generator using the TT network and a forward evaluator based on the fixed VQC.

\textit{Parameter Generation}. In TensorHyper-VQC, all quantum circuit parameters are generated by a classical TT network, thus shifting optimization entirely into the classical domain. Given Gaussian noise input $\textbf{z} \sim \mathcal{N}(0, I)$, the TT network outputs a parameter vector $\hat{\textbf{w}}$ that configures the VQC gates as: 
\begin{equation}
\hat{\textbf{w}} = \text{TT}(\textbf{z}; \{\mathcal{G}_{k}\}_{k=1}^{K}), 
\end{equation}
where each $\mathcal{G}_k$ is a low-rank TT-core, and $K$ represents the number of TT-cores in the TT decomposition. This design ensures that gradient computations and parameter updates are performed solely on the TT-cores, using standard backpropagation in the classical domain. 

\textit{Quantum Forward Pass}. The generated parameters are then injected into a fixed quantum circuit that operates exclusively in inference mode to compute the loss. The quantum circuits, which function exclusively in inference mode to evaluate the loss. The quantum circuits act as a fixed forward-pass evaluator, with no VQC parameters trained. 

This architectural design fundamentally reshapes the landscape of optimization. On the one hand, it insulates gradient computations from barren plateaus and quantum-induced noise instabilities by shifting all trainable parameters to the classical domain. On the other hand, it exploits the TT-induced low-rank structure to achieve intrinsic variance reduction and robustness against measurement noise.

\begin{figure}
\centerline{\epsfig{figure=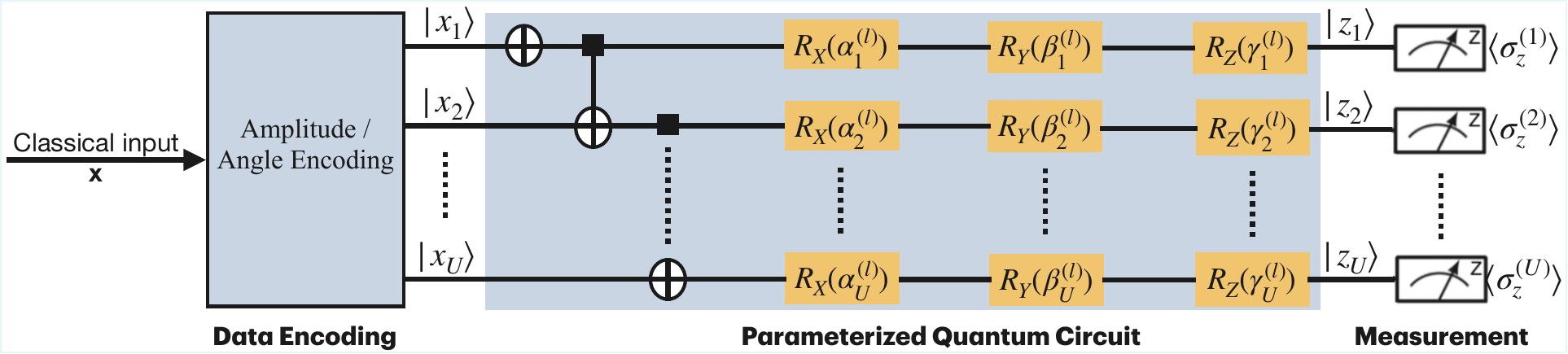, width=130mm}}
\caption{{\it The architecture of variational quantum computing}. The classical input data are encoded as quantum states $\vert x_{1} \rangle$, $\vert x_{2} \rangle$, ..., $\vert x_{U} \rangle$ via data encoding (e.g., angle or amplitude encoding). The shaded region denotes $L$ repeated parameterized quantum circuit block, consisting of trainable rotation gates $R_X(\alpha^{(l)}_u)$, $R_{Y}(\beta^{(l)}_{u})$, $R_{Z}(\gamma^{(l)}_{u})$, and entangling layers across quantum channels. These blocks are stacked to increase circuit depth and expressive capacity. Finally, the circuit is measured in the Pauli-Z basis, yielding expectation values $\langle \sigma_{z}^{(1)} \rangle$, $\langle \sigma_{z}^{(2)} \rangle$, ..., $\langle \sigma_{z}^{(U)} \rangle$ that serve as outputs for optimizaiton.}
\label{fig:vqc}
\end{figure}

Throughout this work, we employ a single, consistent VQC architecture across all experiments and applications. The parameterized quantum circuit block, illustrated in Figure~\ref{fig:vqc}, is fixed in structure: it consists of alternating layers of single-qubit rotations $R_{X}(\alpha^{(l)}_{u})$, $R_{Y}(\beta^{(l)}_{u})$, $R_{Z}(\gamma^{(l)}_{u})$ and entangling gates across $U$ qubits, stacked to achieve the desired circuit depth $L$. By holding the VQC design constant, our results isolate the contribution of the TensorHyper-VQC parameterization strategy, rather than relying on task-specific ansatz engineering. This highlights that the observed gains in trainability, robustness, and scalability stem directly from the TT-based hypernetwork, demonstrating its versatility as a general-purpose parameter generator for VQC. 

Within TensorHyper-VQC, the TT network functions as a classical hypernetwork generator: instead of tuning $\hat{\textbf{w}}$ directly on quantum hardware, we optimize the TT-cores $\{\mathcal{G}_{k}\}_{k=1}^{K}$ entirely in the classical domain. This confers two advantages: (1) gradient with respect to TT-cores occurs in the classical domain, insulating optimization from quantum noise, and (2) the TT-induced low-rank structure yields noise averaging and variance reduction in gradient estimates, as established in our theoretical analysis. 

Finally, a consistent TT hypernetwork architecture is adopted across all applications. The TT ranks and core dimensions are chosen to balance expressivity and generalization, while the same generator design is used across tasks. This uniformity underscores that the improvements in scalability, robustness, and trainability derive from the TT-guided parameterization principle itself rather than from task-specific quantum architectural search.

\subsection{Theoretical Guarantees}

We rigorously provide a theoretical foundation for the advantages of the TensorHyper-VQC framework from four perspectives: representation capability, trainability, generalization performance, and robustness against quantum noise. 

\textit{Approximation Capability.} The approximation capability of TensorHyper-VQC can be analyzed via the approximation error $\epsilon_{\rm app}$, which measures the discrepancy between the expected risk $\mathcal{R}(\cdot)$ of a target function $h^{*}$ and that of the best possible TensorHyper-VQC operator $f_{\boldsymbol{\theta}^{*}}$. This bound follows:
\begin{equation}
\begin{split}
\epsilon_{\rm app} &\le \mathcal{R}(h^{*}) - \mathcal{R}(f_{\boldsymbol{\theta^{*}}}) \\
	&\le K_{\ell} \lVert h^{*}(\textbf{x}) -  f_{\boldsymbol{\theta}^{*}}(\textbf{x}) \rVert_{1} \\
	&\le K_{\ell} \lVert h^{*}(\textbf{x}) - f_{\textbf{w}^{*}}(\textbf{x}) \rVert_{1}  +   K_{\ell} \lVert  f_{\textbf{w}^{*}}(\textbf{x}) - f_{\boldsymbol{\theta}^{*}}(\textbf{x}) \rVert_{1},
\end{split}
\end{equation}
where the expected risk $\mathcal{R}(\cdot)$ is assumed as $K_{\ell}$-Lipschitz continuous, $f_{\textbf{w}^{*}}$ is an optimal VQC operator with parameters $\textbf{w}^{*}$, and $f_{\boldsymbol{\theta}^{*}}$ is the optmal TensorHyper-VQC operator with TT-core parameters $\boldsymbol{\theta}^{*} = \{\mathcal{G}^{*}_{k}\}_{k=1}^{K}$. Given the VQC depth $L$ and qubit count $U$, the term $\lVert h^{*}(\textbf{x}) - f_{\textbf{w}^{*}}(\textbf{x}) \rVert_{1}$ is bounded by the VQC expressivity limit $\mathcal{O}(K_{\ell} e^{-\alpha L}) + \mathcal{O}(\frac{K_{\ell}}{2^{\beta U}})$. The second term $\lVert  f_{\textbf{w}^{*}}(\textbf{x}) - f_{\boldsymbol{\theta}^{*}}(\textbf{x}) \rVert_{1}$ reflects the TT parameterization approximation error, denoted as $\mathcal{O}(K_{\ell} \epsilon_{\rm tt})$. 

For a target weight $\textbf{w}^{*}$ reshaped into order-K with dimensions $(I_1, I_2, ..., I_{K})$, the TT approximation with TT-ranks $\{r_0, r_1, ..., r_{K}\}$ (where $r_0 = r_K = 0$) satisfies: 
\begin{equation}
\epsilon_{\rm tt} = \left\lVert \textbf{w}^{*} - \hat{\textbf{w}} \right\rVert_{1} \le \left(\sum\limits_{k=1}^{K-1} \sum\limits_{j > r_{k}} \sigma_{j}^{2}\left(\mathcal{M}_{k}(\textbf{w}^{*})\right)\right)^{\frac{1}{2}}, 
\end{equation}
where $\mathcal{M}_{k}(\textbf{w}^{*})$ is the mode-$k$ unfolding of $\textbf{w}^{*}$, and $\sigma_{j}(\mathcal{M}_{k}(\textbf{w}^{*}))$ are its singular values. More importantly, the TT approximation error $\epsilon_{\rm tt} = \epsilon_{\rm tt}(r_1, r_2, ..., r_{K-1})$ is a monotonically decreasing function of the TT-ranks, vanishing when each $r_k$ equals the full unfolding rank. Combining these results gives:
\begin{equation}
\epsilon_{\rm app} \le \mathcal{O}\left(K_{\ell} e^{-\alpha L}\right) + \mathcal{O}\left( \frac{K_{\ell}}{2^{\beta U}} \right) + \mathcal{O}\left( K_{\ell} \epsilon_{\rm tt}(r_1, r_2, ..., r_{K-1})	\right),
\end{equation}
which leads to Corollary~\ref{ref:cor1}. 

\begin{corollary}[Rank-Expressivity Trade-off]
\label{ref:cor1}
Let $r = \max_{1\le k\le K-1} r_{k}$ denote the maximum TT-rank. If the singular values of each unfolding $\mathcal{M}_{k}(\textbf{w}^{*})$ decay at least polynomially, i.e., $\sigma_{j}(\mathcal{M}_{k}(\textbf{w}^{*})) = \mathcal{O}(j^{-p})$ for some $p > 1$, then the TT approximation error satisfies:
\begin{equation}
\epsilon_{\rm tt}(r) = \mathcal{O}\left(	 \frac{1}{r^{p-1}} \right). 
\end{equation}
\end{corollary}

Consequently, increasing the TT-ranks reduces the approximation error at a quantifiable rate: low ranks yield compact parameterization with large error. In contrast, higher ranks progressively recover the full expressivity of the VQC. 

\textit{Optimization Dynamics via Neural Tangent Kernel.} Since $\hat{\textbf{w}}$ is not directly parameterized but generated by a TT network and the parameters $\boldsymbol{\theta} = \{\mathcal{G}_{k}\}_{k=1}^{K}$ lie within the TT representation, defining the hierarchical parameterization: 
\begin{equation}
\boldsymbol{\theta} \rightarrow \hat{\textbf{w}}(\boldsymbol{\theta}) \rightarrow f_{\boldsymbol{\theta}}(\textbf{x}; \hat{\textbf{w}}(\boldsymbol{\theta})). 
\end{equation}

We then leverage the NTK theory to analyze the trainability of TensorHyper-VQC. The NTK matrix for our framework, $\mathcal{T}_{\rm tv}$, is determined jointly by the VQC and the TT network as: 
\begin{equation}
\mathcal{T}_{\rm tv} = \nabla_{\boldsymbol{\theta}}f_{\boldsymbol{\theta}}(\textbf{x}) \nabla_{\boldsymbol{\theta}}f_{\boldsymbol{\theta}}(\textbf{x})^{\top} =  \left( \frac{\partial f_{\boldsymbol{\theta}}(\textbf{x})}{\partial \hat{\textbf{w}}} \cdot \frac{\partial \hat{\textbf{w}}}{\partial \boldsymbol{\theta}} \right) \left( \frac{\partial f_{\boldsymbol{\theta}}(\textbf{x}) }{\partial \hat{\textbf{w}}} \cdot \frac{\partial \hat{\textbf{w}}}{\partial \boldsymbol{\theta}} \right)^{\top}. 
\end{equation}

Let the VQC Jacobian as $\textbf{J}_{\textbf{w}} = \frac{\partial f_{\boldsymbol{\theta}}}{\partial \hat{\textbf{w}}}$ and the TT-core Jacobian as $\textbf{J}_{\boldsymbol{\theta}} = \frac{\partial \hat{\textbf{w}}}{\partial \boldsymbol{\theta}}$. Then, we obtain: 
\begin{equation}
\mathcal{T}_{\rm tv} = \textbf{J}_{\hat{\textbf{w}}} \textbf{J}_{\boldsymbol{\theta}} \textbf{J}_{\boldsymbol{\theta}}^{\top} \textbf{J}_{\hat{\textbf{w}}}^{\top}. 
\end{equation}

For any matrices $\textbf{J}_{\hat{\textbf{w}}}$ and $\textbf{J}_{\boldsymbol{\theta}}$, there is
\begin{equation}
\lambda_{\rm min}(\mathcal{T}_{\rm tv}) = \lambda_{\rm min}(\textbf{J}_{\hat{\textbf{w}}} \textbf{J}_{\boldsymbol{\theta}} \textbf{J}_{\boldsymbol{\theta}}^{\top} \textbf{J}_{\hat{\textbf{w}}}^{\top}) \ge \lambda_{\rm min}(\textbf{J}_{\boldsymbol{\theta}} \textbf{J}_{\boldsymbol{\theta}}^{\top}) \lambda_{\rm min}(\textbf{J}_{\hat{\textbf{w}}} \textbf{J}_{\hat{\textbf{w}}}^{\top}), 
\end{equation}
and since a conventional VQC ensures that its minimum NTK eigenvalue is exactly $\lambda_{\rm min}(\textbf{J}_{\hat{\textbf{w}}} \textbf{J}_{\hat{\textbf{w}}}^{\top})$, we obtain: 
\begin{equation}
\lambda_{\rm min}(\mathcal{T}_{\rm vqc}) = \lambda_{\rm min}(\textbf{J}_{\hat{\textbf{w}}} \textbf{J}_{\hat{\textbf{w}}}^{\top}). 
\end{equation}

The TT-core Jacobian satisfies a dispersion property (given the VQC depth $L$, qubit count $U$, and a constant $c$, each TT-core contributes nearly uniformly, so $\textbf{J}_{\boldsymbol{\theta}} \textbf{J}_{\boldsymbol{\theta}}^{\top}$ is well-conditioned with $\lambda_{\rm min}(\textbf{J}_{\boldsymbol{\theta}} \textbf{J}_{\boldsymbol{\theta}}^{\top}) \approx \frac{c}{3UL} > 0$), we dervie: 
\begin{equation}
\lambda_{\rm min}(\mathcal{T}_{\rm tv}) \ge \frac{c}{3UL} \lambda_{\rm min}(\textbf{J}_{\hat{\textbf{w}}} \textbf{J}_{\hat{\textbf{w}}}^{\top}) = \frac{c}{3UL} \lambda_{\rm min}(\mathcal{T}_{\rm vqc}),
\end{equation} 
with a constant $c$ depending on TT-core dispersion. 

This shows that the effective NTK eigenvalue is not smaller and, under mild dispersion assumptions, can be strictly larger, thereby ensuring faster convergence and mitigating vanishing gradients. Accordingly, at epoch $t$, the optimization error satisfies: 
\begin{equation}
\epsilon_{\rm opt}(t) \le C_{0} e^{-\lambda_{\rm min}(\mathcal{T}_{\rm tv})t},
\end{equation}
where $C_0$ depends on the initialized parameters $\boldsymbol{\theta}_{0}$. 

\textit{Generalization Performance.} Around the initialization of parameters $\boldsymbol{\theta}_{0}$, we linearize the operator $f_{\boldsymbol{\theta}}$ as:
\begin{equation}
f_{\boldsymbol{\theta}}(\textbf{x}) \approx f_{\boldsymbol{\theta}_{0}}(\textbf{x}) + \nabla_{\boldsymbol{\theta}}f_{\boldsymbol{\theta}_{0}}(\textbf{x})^{\top}(\boldsymbol{\theta} - \boldsymbol{\theta}_{0}).
\end{equation}

Suppose that the quantum measurement noise is represented by a random vector $\boldsymbol{\tau} \in \mathbb{R}^{3U}$ with mean $\textbf{b} = \operatorname{E}[\boldsymbol{\tau}]$ and covariance matrix $\boldsymbol{\Sigma} = \text{Cov}(\boldsymbol{\tau})$, and given $K_{\ell}$-Lipschitz loss function and dataset $\{(\textbf{x}_{n}, \textbf{y}_{n})\}_{n=1}^{N}$, the generalization gap obeys:
\begin{equation}
\operatorname{E}\left[	\sup\limits_{\boldsymbol{\theta}} \left\vert \mathcal{R}(f_{\boldsymbol{\theta}}) - \hat{\mathcal{R}}(f_{\boldsymbol{\theta}}) \right\vert \right] \le \tilde{O}\left(\sqrt{\frac{\text{Tr}(\mathcal{T}_{\rm tv})}{N^{2}}}\right) + \mathcal{O}\left( \sqrt{\frac{\lambda_{\rm max}(\boldsymbol{\Sigma}) K_{\ell}^{2}}{N}} \right) + \mathcal{O}\left(\frac{\lVert \textbf{b} \rVert_{2} K_{\ell}}{\sqrt{N}}	\right),
\end{equation}
where $\hat{\mathcal{R}}(\cdot)$ denotes the empirical risk, and $\lambda_{\rm max}(\boldsymbol{\Sigma})$ is the maximum eigvenvalue of $\boldsymbol{\Sigma}$. With TT-ranks $\{r_{0}, r_{1}, ..., r_{K}\}$ (where $r_{0} = r_{K} = 0$), we obtain:
\begin{equation}
\text{Tr}(\mathcal{T}_{\rm tv}) \le N C_r\prod\limits_{i=1}^{K} r_{i},
\end{equation}
for some constant $C_r$ (depending on input scaling, circuit, and TT initialization). This result demonstrates that the TT network's structured, low-rank nature helps control the trace of $\mathcal{T}_{\rm tv}$, potentially improving the generalization performance of TensorHyper-VQC, as shown in Corollary~\ref{cor2}.

\begin{corollary}[Expressivity-Generalization Trade-off]
\label{cor2}
Let $r = \max_{i}r_{i}$. Then increasing $r$ reduces the TT approximation error $\epsilon_{\rm tt}(r)$, and hence improves expressivity, but simultaneously enlarges $\text{Tr}(\mathcal{T}_{\rm tv})$, leading to a larger upper bound on the expected generalization gap $\operatorname{E}\left[ \sup_{\boldsymbol{\theta}} \left\vert \mathcal{R}(f_{\boldsymbol{\theta}}) - \hat{\mathcal{R}}(f_{\boldsymbol{\theta}}) \right\vert \right]$. 
\end{corollary}

\textit{Robustness to Quantum Noise.} We finally verify that TensorHyper-VQC offers significant robustness against quantum hardware noise. In the TensorHyper-VQC framework, since we optimize over classical TT-cores $\{\mathcal{G}_{k}\}_{k=1}^{K}$, the quantum circuit no longer introduces vanishing gradients. The gradients are computed via 
\begin{equation}
\frac{\partial \hat{\mathcal{R}}}{\partial \mathcal{G}_{k}} = \frac{\partial \hat{\mathcal{R}}}{\partial \hat{\textbf{w}}} \cdot \frac{\partial \hat{\textbf{w}}}{\partial \mathcal{G}_{k}} = \sum\limits_{u=1}^{3UL} \frac{\partial \hat{\mathcal{R}}}{\partial \hat{w}_{u}} \cdot \frac{\partial \hat{w}_{u}}{\partial \mathcal{G}_{k}} . 
\end{equation}

Since TensorHyper-VQC admits a fully classical training paradigm, $\frac{\partial \hat{\textbf{w}}}{\partial \mathcal{G}_{k}}$ is deterministic and computed classically, while $\frac{\partial \hat{\mathcal{R}}}{\partial \hat{\textbf{w}}}$ is affected by the measurement noise $\boldsymbol{\tau} = [\tau_{1}, \tau_{2}, ..., \tau_{3UL}]^{\top}$, giving: 
\begin{equation}
\frac{\partial \hat{\mathcal{R}}}{\partial \mathcal{G}_{k}} = \left(\frac{\partial \mathcal{R}}{\partial \hat{\textbf{w}}} + \boldsymbol{\tau}\right) \cdot \frac{\partial \hat{\textbf{w}}}{\partial \mathcal{G}_{k}} = \sum\limits_{u=1}^{3UL} \frac{\partial \mathcal{R}}{\partial \hat{w}_{u}} \cdot \frac{\partial \hat{w}_{u}}{\partial \mathcal{G}_{k}} + \sum\limits_{u=1}^{3UL} \tau_{u} \frac{\partial \hat{w}_{u}}{\partial \mathcal{G}_{k}}. 
\end{equation}

Taking the variance of the noisy gradient term $\frac{\partial \hat{\mathcal{R}}}{\partial \mathcal{G}_{k}}$ for the TT-cores, we have: 
\begin{equation}
\text{Var}\left[ \frac{\partial \hat{\mathcal{R}}}{\partial \mathcal{G}_{k}} \right] = \text{Var}\left[ \sum\limits_{u=1}^{3UL} \tau_{u} \frac{\partial \hat{w}_{u}}{\partial \mathcal{G}_{k}} \right]. 
\end{equation} 

We first recall the TT-Jacobian dispersion property used in our baseline derivation, where for the TT-core Jacobian $\textbf{J}_{k} = \frac{\partial \hat{\textbf{w}}}{\partial \mathcal{G}_{k}}$, we have: 
\begin{equation}
\label{eq:disp}
\lVert	 \textbf{J}_{k} \rVert_{2}^{2} \approx \frac{c}{3UL}, \hspace{6mm} c \ll 3UL,
\end{equation}
reflecting the uniform spread of sensitivity across parameters. 

Finally, we analyze the noise robustness of TensorHyper-VQC under three increasingly realistic assumptions about measurement noise: (i) Zero-mean i.i.d. noise (finite-variance or sub-Gaussian); (ii) Zero-mean, non-i.i.d. noise (correlated or heteroskedastic); (iii) Non-zero-mean non-i.i.d. noise (biased measurement). 

We first consider that the measurement noise $\boldsymbol{\tau} = \{\tau_{u}\}_{u=1}^{U}$ is independent, zero-mean, with $\text{Var}(\tau_{u}) = \sigma^{2}$, where no Gaussianity is required, but the condition of sub-Gaussian or finite-variance suffices. This property of independence gives: 
\begin{equation}
\text{Var}\left[ \frac{\partial \hat{\mathcal{R}}}{\partial \mathcal{G}_{k}} \right]  \le \sigma^{2} \lVert \textbf{J}_{k} \rVert_{2}^{2} \lesssim \frac{c\sigma^{2}}{3UL}. 
\end{equation}
Thus, we obtain the $\mathcal{O}(\frac{1}{UL})$ variance reduction of TT-core gradients under the zero-mean, i.i.d. noisy condition. 

If the measurement noise $\boldsymbol{\tau}$ is assumed to be zero-mean with unknown covariance $\boldsymbol{\Sigma} \in \mathbb{R}^{3UL \times 3UL}$, where entries could be unequal and correlated, which leads to
\begin{equation}
\text{Var}\left[ \frac{\partial \hat{\mathcal{R}}}{\partial \mathcal{G}_{k}} \right] \le \lambda_{\rm max}(\boldsymbol{\Sigma}) \lVert \textbf{J}_{k} \rVert_{2}^{2} \lesssim \lambda_{\rm max}(\boldsymbol{\Sigma})\frac{c}{3UL}. 
\end{equation}

Hence, the $\mathcal{O}(\frac{1}{UL})$ decay persists under arbitrary zero-mean correlations, where only the constant changes via $\lambda_{\rm max}$. In particular, if $\boldsymbol{\Sigma}$ is diagonally dominant (heteroskedastic readout) with row sums bounded by $B$, we also obtain: 
\begin{equation}
\text{Var}\left[ \frac{\partial \hat{\mathcal{R}}}{\partial \mathcal{G}_{k}} \right] \lesssim \frac{c B}{3UL}. 
\end{equation}

Furthermore, if we assume that the quantum measurement noise is $\boldsymbol{\tau} \in \mathbb{R}^{3UL}$ with mean $\textbf{b}:= \operatorname{E}[\boldsymbol{\tau}]$ and covariance $\boldsymbol{\Sigma} = \text{Cov}(\boldsymbol{\tau})$. With $\textbf{J}_{k} := \frac{\partial \hat{\textbf{w}}}{\partial \mathcal{G}_{k}}$, the TT-core gradient estimator is 
\begin{equation}
\hat{\textbf{g}}_{k} = \frac{\partial \hat{\mathcal{R}}}{\partial \mathcal{G}_{k}} = \textbf{g}_{k} + \textbf{J}_{k}^{\top} \boldsymbol{\tau}, \hspace{4mm} \textbf{g}_{k} := \frac{\partial \mathcal{R}}{\partial \mathcal{G}_{k}}. 
\end{equation}

Hence, we obtain the bias and variance terms as: 
\begin{equation}
\operatorname{E}[\hat{\textbf{g}}_{k}] - \textbf{g}_{k} = \textbf{J}_{k}^{\top} \textbf{b}, \hspace{4mm} \text{Var}(\hat{\textbf{g}}_{k}) = \textbf{J}_{k}^{\top} \boldsymbol{\Sigma} \textbf{J}_{k},
\end{equation}
which separately leads to their upper bounds as: 
\begin{equation}
\left\lVert \operatorname{E}[\hat{\textbf{g}}_k] - \textbf{g}_{k} \right\rVert_{2} \le \lVert \textbf{J}_k \rVert_{2} \lVert \textbf{b} \rVert_{2} \le \sqrt{\frac{c}{3UL}} \lVert \textbf{b} \rVert_{2}, 
\end{equation}
\begin{equation}
\text{Var}(\hat{\textbf{g}}_k) \le \lambda_{\rm max}(\boldsymbol{\Sigma}) \frac{c}{3UL}. 
\end{equation}

In particular, the variance term keeps the $\mathcal{O}(\frac{1}{UL})$ improvement due to the TT Jacobian spreading, while the bias term also scales with $\mathcal{O}(\frac{1}{UL})$ provided $\lVert \textbf{b} \rVert_{2}$ does not grow with $U$.

By contrast, a standard VQC parameterized directly by gate weights $\textbf{w}$, the gradient estimator takes the form of
\begin{equation}
\hat{\textbf{g}}_{\textbf{w}} = \frac{\partial \hat{\mathcal{R}}}{\partial \textbf{w}} = \frac{\partial \mathcal{R}}{\partial \textbf{w}} + \boldsymbol{\tau}_{\textbf{w}},
\end{equation}
where $\boldsymbol{\tau}_{\textbf{w}}$ denotes the measurement noise associated with that parameter. 

In this case, the noise characteristics directly propagate to the gradients with two noisy conditions: (i) i.i.d. finite-variance noise, and (ii) non-i.i.d. or biased noise. 

Under the case of i.i.d. finite-variance noise, we obtain: 
\begin{equation}
\text{Var}(\hat{\textbf{g}}_{\textbf{w}}) = \sigma^{2}, 
\end{equation}
which remains constant, independent of the number of qubits. 

As for the case of non-i.i.d. or biased noise, we derive:
\begin{equation}
\text{Var}(\hat{\textbf{g}}_{\textbf{w}}) \approx \lambda_{\rm max}(\boldsymbol{\Sigma}), \hspace{4mm} \operatorname{E}[\hat{\textbf{g}}_{\textbf{w}}] - \textbf{g}_{\textbf{w}} = \textbf{b}, 
\end{equation}
where both variance and bias persist without attenuation as the system scales. 

Thus, unlike TensorHyper-VQC, the noise impact in conventional VQC does not diminish with qubit count. This makes training increasingly fragile in larger systems, as noise accumulates rather than averaging out. In contrast, the TT-guided hypernetwork spreads sensitivity across TT-cores, yielding a natural noise-robust averaging effect: gradient variance scales as $\mathcal{O}(\frac{1}{UL})$ and bias contributions are suppressed by $(UL)^{-\frac{1}{2}}$.

\subsection{Implementation of Gradient Descent on Quantum Hardware}
To demonstrate the practicality of TensorHyper-VQC, we outline how gradient-based optimization can be implemented on a real quantum device. The hybrid training loop integrates classical TT-core updates with quantum circuit evaluations. 

For a given input $\textbf{x}$, the TT hypernetwork with parameters $\boldsymbol{\theta} = \{\mathcal{G}_{k}\}_{k=1}^{K}$ generates the variational parameters $\hat{\textbf{w}}(\boldsymbol{\theta})$. These parameters are loaded into the quantum circuit ansatz $\mathcal{U}(\hat{\textbf{w}}(\boldsymbol{\theta}))$, and the circuit expectation value is estimated via repeated measurements as: 
\begin{equation}
f_{\boldsymbol{\theta}}(\textbf{x}) = \langle 0 \vert \mathcal{U}^{\dagger} (\hat{\textbf{w}}(\boldsymbol{\theta})) \left( \sum\limits_{u=1}^{U} \sigma_{z}^{(u)}	\right) \mathcal{U} (\hat{\textbf{w}}(\boldsymbol{\theta})) \vert 0 \rangle. 
\end{equation}

On real hardware, analytic gradients with respect to each variational parameter can be obtained using the parameter-shift rule. For a parameter $\hat{w}_{u}$, the derivative is computed as: 
\begin{equation}
\frac{\partial f_{\boldsymbol{\theta}}(\textbf{x})}{\partial \hat{w}_{u}} = \frac{1}{2} \left( f_{\boldsymbol{\theta}}\left(\textbf{x}; \hat{w}_{u} + \frac{\pi}{2}\right) - f_{\boldsymbol{\theta}}\left(\textbf{x}; \hat{w}_{u} - \frac{\pi}{2}\right) \right),
\end{equation}
which requires two additional circuit evaluations per parameter. 

Since TensorHyper-VQC decouples trainable TT-cores from quantum evaluations, gradients with respect to TT parameters follow naturally from the chain rule as: 
\begin{equation}
\frac{\partial f_{\boldsymbol{\theta}}(\textbf{x})}{\partial \mathcal{G}_{k}} = \sum\limits_{u=1}^{3UL} \frac{\partial f_{\boldsymbol{\theta}}(\textbf{x})}{\partial \hat{w}_{u}} \cdot \frac{\partial \hat{w}_{u}}{\partial \mathcal{G}_k},
\end{equation}
where the Jacobian $\frac{\partial \hat{w}_{u}}{\partial \mathcal{G}_k}$ is computed entirely in the classical domain using CPU or GPU resources. 

This hybrid procedure enables TensorHyper-VQC to perform gradient-based optimization on real quantum processors efficiently. The TT network handles all differentiable computation classically, while the quantum hardware is used solely for forward-pass evaluations and measurement sampling, ensuring scalability and robustness against quantum noise.

\subsection{Application to Quantum Machine Learning: Quantum Dot Classification}

We first apply TensorHyper-VQC to a binary classification task on charge-stability diagrams from semiconductor quantum dots~\cite{tang2015storage, kalantre2019machine, gualtieri2025qdsim} to validate our theoretical framework. The objective is to classify diagrams into single-dot (Label 0) or double-dot (Label 1) systems, a crucial task for characterizing and controlling quantum devices. The experimental dataset comprises $50 \times 50$-pixel images associated with quantum dot diagrams, including $2,000$ diagrams generated under realistic device-level noise conditions. We randomly partition the diagrams into $1,800$ training and $200$ test samples to evaluate the models' generalization performance. 

\begin{figure}
\centerline{\epsfig{figure=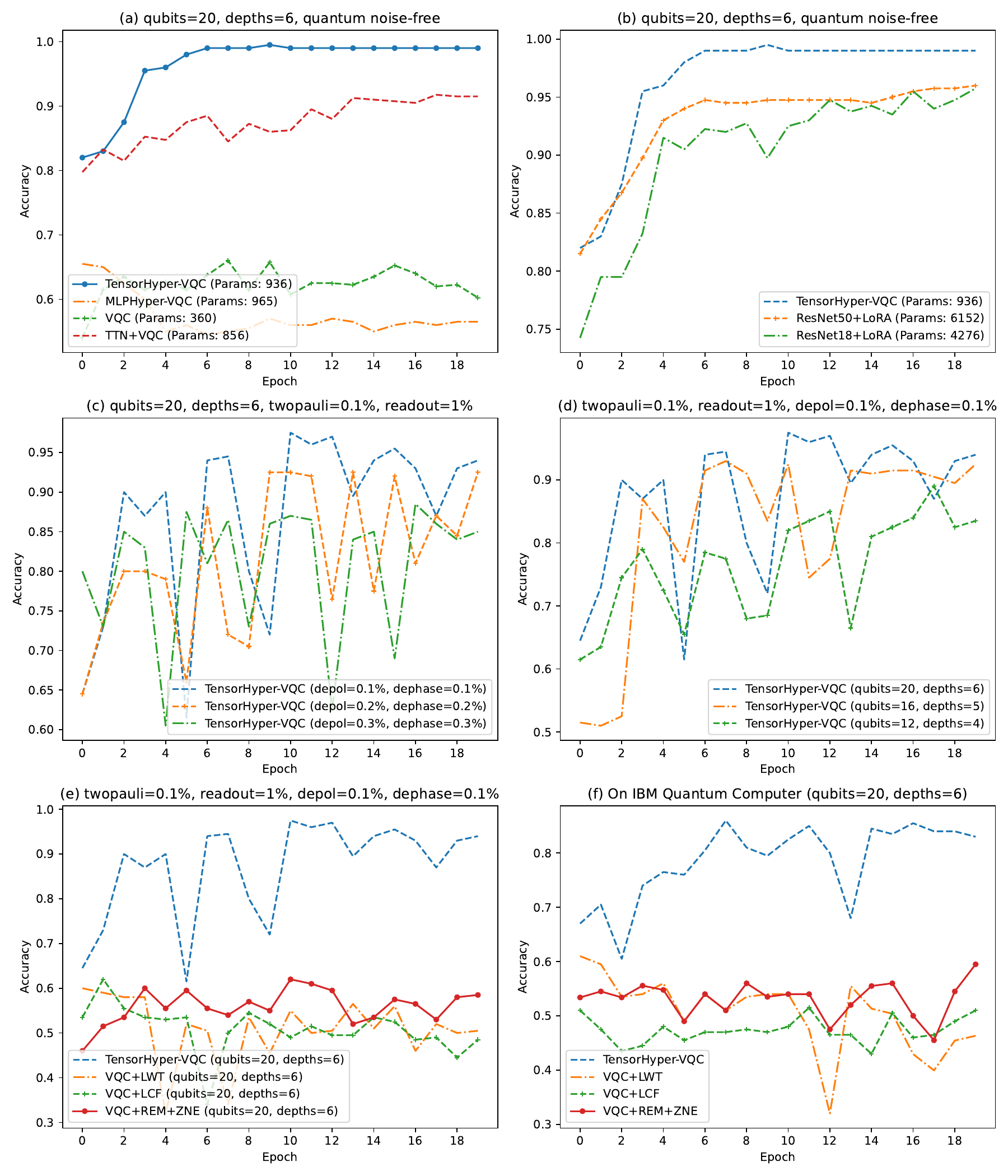, width=150mm}}
\caption{{\it Comprehensive performance evaluation of TensorHyper-VQC for quantum-dot classification, measured by accuracy on the test dataset}. (a) Comparison with standard VQC, MLPHyper-VQC, and TTN+VQC under noiseless conditions, showing faster convergence and higher accuracy; (b) Comparison with classical baselines (ResNet18+LoRA and ResNet50+LoRA), where TensorHyper-VQC achieves superior accuracy with far fewer parameters; (c) Robustness under increasing depolarizing and dephasing noise, maintaining stable performance; (d) Scalability across different qubit counts, with graceful accuracy degradation as system size increases; (e) Comparative analysis under realistic multi-noise conditions, where TensorHyper-VQC outperforms VQC enhanced with LWT, LCF, and REM + ZNE, demonstrating improved trainability and noise resilience; (f) Experimental validation on a 156-qubit IBM quantum processor (Heron r2), confirming its performance advantages and stable convergence without additional mitigation overhead.}
\label{fig:qd_res}
\end{figure}

The proposed TensorHyper-VQC architecture integrates a classical TT network as a parameter generator for VQC. The TT network is configured with input dimensions $[5, 10, 5, 10]$, output dimensions $[4, 2, 3, 9]$, and TT-ranks $[1, 2, 2, 2, 1]$, resulting in a compact representation with only $576$ trainable parameters (including bias). The TT network maps vectorized quantum-dot features to all rotation angles required by the VQC, thereby parameterizing $20$ qubits across $6$ variational layers (360 quantum-gate parameters). All models are trained using the cross-entropy loss~\cite{mao2023cross} based on the Adam optimizer~\cite{adam:2014} with a fixed learning rate $3\times 10^{-2}$. We use a fixed random seed across all experiments to ensure reproducibility and fairness, and we initialize parameters from a normal distribution with mean 0 and variance 1. 

To comprehensively benchmark performance, we compare TensorHyper-VQC against multiple baselines, including (i) a standard VQC with randomly initialized parameters, (ii) MLPHyper-VQC, where a multilayer perceptron replaces the TT network as the parameter generator, (iii) TTN+VQC, a hybrid model combining a TT encoder and VQC~\cite{qi2023theoretical, chen2021end}, and (iv) ResNet18~\cite{szegedy2017inception} and ResNet50~\cite{wen2020transfer} enhanced with low-rank adaptation (LoRA)~\cite{hu2022lora, liu2024dora} for classical comparison. Furthermore, we conduct robustness analyses by injecting a variety of realistic noise models into the quantum circuits, including depolarizing, dephasing, two-Pauli, and readout noise, with error probabilities ranging from $0.1\%$ to $0.3\%$, simulating typical NISQ-era conditions. 

To further assess resilience to noise, we incorporate standard barren-plateau mitigation and error-reduction techniques, including layer-wise training (LWT)~\cite{skolik2021layerwise}, which progressively optimizes circuit depth to maintain gradient signal, and local cost function (LCF) design~\cite{cerezo2021cost}, which mitigates gradient vanishing by focusing on local observables. In addition, we employ standard quantum error mitigation approaches, including zero-noise extrapolation (ZNE)~\cite{he2020zero}, readout error mitigation (REM)~\cite{smith2021qubit}, and their combination (REM+ZNE). We evaluate the performance of all these strategies on both simulated and physical backends. Finally, we validate our approach on a 156-qubit IBM Quantum processor (Heron r2), demonstrating the deployability of TensorHyper-VQC on real hardware and its robustness to device-level imperfections, including decoherence, crosstalk, and measurement errors.

\textit{Performance in the Quantum Noise-Free Regime}. We first benchmark TensorHyper-VQC against standard VQC, MLPHyper-VQC, and the hybrid TTN+VQC model in a noise-free setting, as shown in Fig.~\ref{fig:qd_res}(a). TensorHyper-VQC rapidly converges to near-perfect accuracy ($99.5\%$) within $10$ epochs, significantly outperforming both standard VQC ($65.7\%$) and TTN+VQC ($91.7\%$), as well as the MLPHyper-VQC baseline (56.5\%). Compared with deep classical models (ResNet18+LoRA and ResNet50+LoRA), TensorHyper-VQC also achieves higher accuracy with $4-6x$ fewer parameters, illustrated in Fig.~\ref{fig:qd_res}(b). These results empirically confirm that the TensorHyper-VQC architecture offers a distinct advantage in expressivity and optimization efficiency compared with conventional quantum, hybrid, and classical approaches. 

\textit{Robustness Under Multi-Noise Conditions}. We next evaluate the robustness under realistic quantum noise models~\cite{cross2019validating, resch2021benchmarking}, including depolarizing, dephasing, two-Pauli, and readout noise channels, as illustrated in Fig.~\ref{fig:qd_res}(c). Even as noise levels as high as $0.3\%$, TensorHyper-VQC maintains strong performance ($\ge 85\%$), significantly outperforming noise-agnostic VQC baselines. This robustness arises from the TT-based parameterization, which smooths the optimization landscape and enhances noise tolerance. Furthermore, as shown in Fig.~\ref{fig:qd_res}(d), TenosrHyper-VQC exhibits graceful degradation when scaling from $20$ qubits down to $16$ or $12$ qubits, demonstrating robustness to hardware resource constraints while retaining competitive performance. 

\textit{Comparison with Quantum Error Mitigation}. We also compare TensorHyper-VQC with standard quantum error mitigation techniques, including ZNE, REM, and their combination as shown in Fig.~\ref{fig:qd_res}(e). Remarkably, TensorHyper-VQC achieves a final accuracy of $94.0\%$, outperforming VQC enhanced with ZNE+REM ($55.7\%$) by nearly $40$ percentage points. This highlights that our tensor-structured parameterization intrinsically mitigates noise by reducing parameter variance and constraining quantum state evolution, often outperforming post-hoc mitigation methods that require additional circuit executions and calibration overhead. 

\textit{Real-Device Validation on IBM Quantum Hardware}. We deploy TensorHyper-VQC on a $156$-qubit IBM quantum processor (Heron r2) to assess real-hardware performance as shown in Fig.~\ref{fig:qd_res}(f). Despite inevitable device-level imperfections such as decoherence, gate errors, crosstalk, and readout noise, TensorHyper-VQC achieves a peak accuracy of $86.0\%$ and a final accuracy of $83.0\%$. In contrast, mitigation-enhanced baselines plateau around $50\%$-$53\%$, demonstrating the intrinsic hardware-level noise resilience of the proposed TensorHyper-VQC. These results confirm that TensorHyper-VQC not only excels in simulation but also scales effectively to real NISQ hardware. 

\textit{Ablation Study on TT-Rank Selection}. To further investigate the trade-offs predicted by our theoretical framework, we conduct an ablation study by varying the maximum TT-rank from 2 to 5. The results, summarized in Table~\ref{tab:tt_rank_ablation}, show that increasing TT-rank leads to a monotonic rise in the number of trainable parameters (from 936 to 1146) and enhanced expressivity, as reflected in higher training accuracy (peaking at $96.44\%$). However, the relationship with generalization is non-monotonic: the compact configuration $[1, 2, 2, 2, 1]$ achieves the best test accuracy ($94.00\%$), whereas larger TT-ranks result in degraded generalization (down to $90.50\%$ and $89.50\%$) due to overfitting. This observation reflects the classical bias–variance trade-off, where excessive representational capacity amplifies sensitivity to noise.

\begin{table}[ht]
\centering
\caption{Ablation study on TT-rank selection under the TensorHyper-VQC's configuration and noisy quantum setting described in Fig. (e).}
\label{tab:tt_rank_ablation}
\begin{tabular}{c c c c}
\hline
\textbf{TT-ranks} & \textbf{\# Params} & \textbf{Train Accuracy ($\%$)} & \textbf{Test Accuracy (\%)} \\
\hline
$[1,2,2,2,1]$ 	& 936 	& 93.39 		& 94.00 \\
$[1,2,3,2,1]$ 	& 1006 	& 91.89 		& 90.50 \\
$[1,2,4,2,1]$ 	& 1076 	& 89.72 		& 93.50 \\
$[1,2,5,2,1]$ 	& 1146 	& 96.44 		& 89.50 \\
\hline
\end{tabular}
\end{table}

In summary, across all evaluation scenarios, including noise-free training, multi-noise robustness tests, error mitigation comparisons, real-hardware experiments, and TT-rank ablations, TensorHyper-VQC consistently demonstrates superior accuracy, trainability, noise resilience, and parameter efficiency. These results validate the proposed TT-guided hypernetwork as a scalable, noise-tolerant solution for near-term quantum machine learning and hardware-constrained quantum applications.

\subsection{Application to Combinatorial Optimization: Max-Cut}

We next evaluate the proposed TensorHyper-VQC architecture on a paradigmatic combinatorial optimization problem, which is the Max-Cut problem, formulated within the classical quantum approximate optimization algorithm (QAOA) framework~\cite{wang2018quantum}. The Max-Cut problem seeks a bipartition of a graph that maximizes the number of edges connecting the two partitions. Formally, for a graph $G = (V, E)$ with vertex set $V = V_1 \cup V_2$, the objective is to maximize the number (or total weight) of edges between $V_1$ and $V_2$.

With Pauli-Z operators $\sigma_{z}^{(i)}$, $\sigma_{z}^{(j)}$ on qubit $i$ and $j$ associated with the cost $c_{ij}$, the Max-Cut cost function is encoded as: 
\begin{equation}
\mathcal{H} = \sum\limits_{(i,j) \in E} c_{ij} \frac{1 - \sigma_{z}^{(i)}\sigma_{z}^{(j)}}{2}. 
\end{equation}

In our experiments, TensorHyper-VQC employs a compact TT network to generate all variational circuit parameters required by QAOA across different graph instances. The TT network is constructed with an input dimension of $10$, output dimensions $[1, 1]$, and TT-ranks $[1, 4, 1]$, resulting in a lightweight parameterization with only $41$ trainable parameters (including bias). We benchmarked the approach on 10 randomly generated 20-qubit Erdos-Renyi graphs~\cite{engel2004large}, in which each edge is included independently with probability $0.5$. 

To establish comparative baselines, we analyze three settings: a classical QAOA, optimized purely by classical methods; a classical QAOA (ZNE+REM), where the training of all circuit parameters is augmented with ZNE and REM; and our TensorHyper-VQC-enhanced QAOA, in which TT cores serve as a hierarchical hypernetwork that efficiently generates quantum-gate angles. For each configuration, the objective is to optimize the expectation value of the Max-Cut Hamiltonian, $\langle \mathcal{H} \rangle$, averaged over $10$ graph realizations. 

\textit{Performance Without Quantum Noise}. As shown in Table~\ref{tab:tab1}, TensorHyper-VQC significantly outperforms classical QAOA across all graph instances in the noiseless simulation. The final expectation value of the Max-Cut Hamiltonian $\langle \mathcal{H} \rangle$ improves from an average of $25.97$ for classical QAOA to $30.21$ for TensorHyper-VQC, yielding an average gain of $+4.24$ $(+16.34\%)$. This confirms that the TT-based parameterization enhances trainability and more effectively explores the optimization landscape than traditional gradient-based QAOA. 

\begin{table}[t]
\centering
\caption{Performance comparison between classical QAOA and the proposed TensorHyper-VQC for solving the Max-Cut problem across 10 randomly generated 20-qubit Erdős–Rényi graphs in a noiseless simulation environment (20-Qubit Graphs).}
\begin{tabular}{cccc}
\hline
\textbf{Graph} & \textbf{Classical QAOA} & \textbf{TensorHyper-VQC} & \textbf{Improvement}  \\
\hline
Graph 01 		& 27.22 		& 31.482 		& +4.26\ \,(\,+15.66\%) \\
Graph 02 		& 26.68 		& 31.139 		& +4.46\ \,(\,+16.71\%) \\
Graph 03 		& 24.29 		& 28.633 		& +4.34\ \,(\,+17.88\%) \\
Graph 04 		& 24.80 		& 29.160 		& +4.36\ \,(\,+17.58\%) \\
Graph 05 		& 24.34 		& 28.012 		& +3.67\ \,(\,+15.09\%) \\
Graph 06 		& 26.88 		& 31.015		& +4.14\ \,(\,+15.39\%) \\
Graph 07 		& 27.22 		& 31.613		& +4.39\ \,(\,+16.14\%) \\
Graph 08 		& 24.64 		& 29.227 		& +4.59\ \,(\,+18.62\%) \\
Graph 09 		& 26.82 		& 30.599 		& +3.78\ \,(\,+14.09\%) \\
Graph 10 		& 26.82 		& 31.248 		& +4.43\ \,(\,+16.51\%) \\
\hline
\textbf{Average} & \textbf{25.97} & \textbf{30.21} & \textbf{+4.24\ \,(\,+16.34\%)}  \\
\hline
\end{tabular}
\label{tab:tab1}
\end{table}

\textit{Comparison Against Noise-Mitigation-Enhanced QAOA}. To assess performance under realistic conditions, we further compare TensorHyper-VQC with classical QAOA enhanced by ZNE and REM, two widely used quantum error-mitigation techniques. The results in Table~\ref{tab:tab2} demonstrate that TensorHyper-VQC still outperforms QAOA (ZNE+REM), with an average improvement of $+4.05$ (or $+16.09\%$). This shows that TensorHyper-VQC achieves competitive or better results without relying on quantum noise mitigation techniques, underscoring its inherent noise resilience and robust parameterization. 

\begin{table}[t]
\centering
\caption{Comparing Classical QAOA enhanced with ZNE+REM versus TensorHyper-VQC on MaxCut over 10 random graph instances (20-Qubit Graphs). The noisy simulation incorporates realistic NISQ-level noise, including depolarizing noise ($0.1\%$), dephasing noise ($0.1\%$), two-qubit Pauli errors ($0.1\%$), and readout error ($1\%$).}
\label{tab:tab2}
\begin{tabular}{lccc}
\hline
\textbf{Graph} & \textbf{QAOA (ZNE+REM)} & \textbf{TensorHyper-VQC} & \textbf{Improvement} \\
\hline
Graph 01 		& 28.93 	& 33.034 		& +4.10\ \,(\,+14.19\%) \\
Graph 02 		& 31.79	& 35.366 		& +3.58\ \,(\,+11.25\%) \\
Graph 03 		& 19.88 	& 24.223 		& +4.34\ \,(\,+21.85\%) \\
Graph 04 		& 21.80 	& 26.032 		& +4.23\ \,(\,+19.41\%) \\
Graph 05 		& 23.92 	& 28.434 		& +4.51\ \,(\,+18.87\%) \\
Graph 06 		& 25.34 	& 29.238 		& +3.90\ \,(\,+15.39\%) \\
Graph 07 		& 27.88 	& 31.905 		& +4.03\ \,(\,+14.44\%) \\
Graph 08 		& 25.99 	& 29.769 		& +3.78\ \,(\,+14.54\%) \\
Graph 09 		& 20.77 	& 24.573 		& +3.80\ \,(\,+18.31\%) \\
Graph 10 		& 25.34 	& 29.553 		& +4.21\ \,(\,+16.62\%) \\
\hline
\textbf{Average} & \textbf{25.16} & \textbf{29.21} & \textbf{+4.05\ \,(\,+16.09\%)}  \\
\hline
\end{tabular}
\end{table}

\textit{Real-Hardware Evaluation on IBM Quantum Devices}. We finally deploy TensorHyper-VQC and noise-mitigated classical QAOA on a $156$-qubit IBM quantum processor (Heron r2) to evaluate real-device performance under hardware noise. As summarized in Table~\ref{tab:tab3}, TensorHyper-VQC consistently outperforms QAOA (ZNE + REM) across all test graphs. The average Max-Cut value increases from $24.66$ to $28.89$, achieving an improvement of $+4.23$ ($+17.16\%$). 

\begin{table}[t]
\centering
\caption{Experimental results on a $156$-qubit IBM quantum processor (Heron r2) comparing Classical QAOA with ZNE+REM noise mitigation and TensorHyper-VQC on the MaxCut problem over 10 random graph instances ($20$-Qubit Graphs).}
\label{tab:tab3}
\begin{tabular}{lccc}
\hline
\textbf{Graph} & \textbf{QAOA (ZNE+REM)} 	& \textbf{TensorHyper-VQC} 	& \textbf{Improvement} \\
\hline
Graph 01 		& 22.80 	& 27.248 		& +4.45\ \,(\,+19.51\%) 	\\
Graph 02 		& 24.11 	& 28.435 		& +4.33\ \,(\,+17.94\%) 	\\
Graph 03 		& 22.64 	& 27.305 		& +4.67\ \,(\,+20.60\%) 	\\
Graph 04 		& 23.68 	& 28.120 		& +4.44\ \,(\,+18.76\%) 	\\
Graph 05 		& 30.18 	& 33.950 		& +3.77\ \,(\,+12.49\%) 	\\
Graph 06 		& 28.00 	& 32.025 		& +4.03\ \,(\,+14.38\%) 	\\
Graph 07 		& 21.95 	& 26.069 		& +4.12\ \,(\,+18.78\%) 	\\
Graph 08 		& 21.71 	& 26.089 		& +4.38\ \,(\,+20.17\%) 	\\
Graph 09 		& 29.20 	& 33.466 		& +4.27\ \,(\,+14.60\%) 	\\
Graph 10 		& 22.36 	& 26.234 		& +3.87\ \,(\,+17.32\%) 	\\
\hline
\textbf{Average} & \textbf{24.66} & \textbf{28.89} & \textbf{+4.23\ \,(\,+17.16\%)}	\\
\hline
\end{tabular}
\end{table}

In summary, these results clearly demonstrate the advantage of TensorHyper-VQC for combinatorial optimization tasks. Even without noise mitigation, it consistently surpasses both classical QAOA and ZNE+REM-enhanced variants, highlighting its ability to generate noise-resilient variational parameters and efficiently explore the solution space. Its superior performance on real IBM hardware validates its practical applicability for NISQ-era quantum optimization.

\subsection{Application to Quantum Simulation: Molecular Hamiltonians}

To further demonstrate the practical utility of the proposed TensorHyper-VQC framework, we evaluated its performance on a canonical quantum simulation task: the ground-state energy estimation of the lithium hydride (LiH) molecule. Quantum chemistry simulations are widely regarded as among the most compelling near-term applications of variational quantum algorithms. It thus serves as a stringent benchmark for assessing the scalability, expressivity, and robustness of variational architectures. In this study, we considered a 4-qubit reduced Hamiltonian of LiH, obtained by freezing core orbitals and truncating high-lying virtual orbitals on an STO-3G basis. This reduced model preserves the essential many-body correlations while enabling tractable simulations on current quantum hardware.

We configured TensorHyper-VQC using a TT hypernetwork to generate the variational ansatz parameters. The ansatz consists of $L=2$ layers of parameterized single-qubit rotations ($R_{X}, R_{Y}, R_{Z}$) followed by a ring of CNOT entanglers, resulting in a 24-parameter variational circuit in the conventional VQE setting. By contrast, the TT hypernetwork compactly represents this parameter vector using input dimensions $[4, 6]$ and TT-ransk $[1, 2, 1]$, thereby reducing the number of trainable parameters to only $9$ without compromising expressivity. The TT parameters were optimized classically using the COBYLA algorithm~\cite{regis2011stochastic}  to minimize the ground-state energy of the LiH Hamiltonian.

\textit{Performance in Ideal and Noisy Simulation Environments}. Table~\ref{tab:lih_results1} summarizes the simulation results under both idealized (noise-free) and realistic (noisy) conditions. In the noiseless regime, TensorHyper-VQC achieved a ground-state energy of $-7.861844$ Ha, corresponding to a deviation of only $+0.000285$ Ha from the full configuration interaction (FCI) reference value of $-7.862129$ Ha. This represents more than an order-of-magnitude improvement in accuracy over a conventional VQE with 24 parameters ($+0.003459$ Ha), while requiring $2.7\times$ fewer parameters. Under NISQ-relevant noise, including depolarizing and dephasing noise ($0.1\%$), two-qubit Pauli errors ($0.1\%$), and readout errors ($1\%$), TensorHyper-VQC maintained strong performance, yielding a final energy error of $+0.009849$ Ha. This is in stark contrast to the classical VQE with ZNE and REM, which exhibited a significantly larger deviation of $+0.041711$ Ha. These results indicate that TT-based hyperparameterization substantially enhances noise resilience by constraining the optimization landscape to a low-rank manifold, thereby mitigating the accumulation of stochastic errors.

\begin{table}[t]
\centering
\caption{Quantum Simulation Results of the LiH Molecule Hamiltonians with and without quantum noise. The noisy conditions include depolarizing noise ($0.1\%$), dephasing noise ($0.1\%$), two-qubit Pauli errors ($0.1\%$), and readout error ($1\%$).} 
\vspace{0.5em}
\begin{tabular}{c|c|c|c}
\hline
\textbf{Method} & \textbf{Params} & \textbf{Energy (Ha)} & \textbf{Error vs.\ FCI (Ha)} 	\\
\hline
Exact (FCI)          & ---  	& $-7.862129$ 		& ---           	   		 \\
\hline
Classical VQE     & 24     	& $-7.858670$ 		& $+0.003459$    		 \\
TensorHyper-VQC       & 9     	& $-7.861844$ 		& $+0.000285$    		 \\
\hline
VQE (ZNE+REM with noise)   & 24      & $-7.820418$ & $+0.041711$     	\\
TensorHyper-VQC (with noise)     & 9        & $-7.852280$ & $+0.009849$      \\
\hline
\end{tabular}
\label{tab:lih_results1}
\end{table}

\textit{Validation on Real Quantum Hardware}. To assess the framework’s robustness under practical hardware constraints, we further implemented the LiH simulation on an IBM Quantum $156$-qubit processor (Heron r2). As shown in Table~\ref{tab:lih_results2}, TensorHyper-VQC once again outperformed the baseline. The experimentally measured ground-state energy was $-7.835377$ ($+0.026752$ Ha error), compared with $-7.809882$ Ha ($+0.052247$ Ha error) achieved by the conventional VQE with ZNE+REM. The significant reduction in error on real hardware underscores the advantage of TT-guided parameterization in mitigating the effects of noise, decoherence, and device imperfections.

\begin{table}[t]
\centering
\caption{Quantum Simulation Results of the LiH Molecule Hamiltonians on a $156$-qubit IBM Quantum Computer (Heron r2).}
\vspace{0.5em}
\begin{tabular}{c|c|c|c}
\hline
\textbf{Method} & \textbf{Params} & \textbf{Energy (Ha)} & \textbf{Error vs.\ FCI (Ha)} 	\\
\hline
Exact (FCI)          & ---  	& $-7.862129$ 		& ---           	   		 \\
\hline
VQE (ZNE+REM)  	    & 24      	& $-7.809882$ & $+0.052247$     	\\
TensorHyper-VQC    	    & 9         	& $-7.835377$ & $+0.026752$      	\\
\hline
\end{tabular}
\label{tab:lih_results2}
\end{table}

Overall, these findings demonstrate that TensorHyper-VQC attains near-chemical-accuracy ground-state energy estimates with dramatically fewer parameters and simultaneously exhibits intrinsic robustness to quantum noise and hardware imperfections. The structured low-rank parameterization yields ansätze that are both expressive and less sensitive to perturbations, leading to superior performance in both simulated and experimental quantum-simulation settings.

\section{Discussion}
In this work, we introduce TensorHyper-VQC, a novel tensor-train-guided hypernetwork architecture designed to address two fundamental challenges in VQC: barren plateaus and susceptibility to quantum noise. By strategically decoupling parameter optimization from quantum hardware and delegating it to a classical TT network, TensorHyper-VQC substantially enhances the scalability, robustness, and practicality of variational quantum algorithms. This classical-quantum decoupling reshapes the optimization landscape, improving gradient flow and mitigating the detrimental effects of noise on parameter updates. 

Conventional VQCs often suffer from vanishing gradients and noise-induced instability, limiting their expressivity and applicability to realistic quantum devices. Numerous recent studies have sought to address these challenges through advanced parameterization strategies~\cite{larocca2023theory} and structural optimization~\cite {holmes2022connecting}. Low-rank tensor decompositions, in particular, have emerged as a practical approach to reduce parameter dimensionality and enhance generalization in both classical and quantum settings. Building on these advances, TensorHyper-VQC leverages an entirely classical TT hypernetwork to generate variational parameters, thereby combining the representational power of tensor networks with the trainability benefits of low-rank optimization.

This architectural shift yields several advantages. First, by transferring all optimization tasks to the classical domain, TensorHyper-VQC mitigates the stochasticity introduced by quantum measurement, yielding more stable and efficient training. Second, the inherent low-rank structure of the TT parameterization introduces beneficial inductive biases, such as noise averaging and variance reduction across TT cores, which directly contribute to noise resilience. Our theoretical analyses, based on NTK theory and variance reduction, further substantiate these benefits by demonstrating that the variance of quantum noise in TT-core gradients decreases with increasing qubit count—an effect not exploited by conventional VQCs.

Empirically, TensorHyper-VQC consistently outperforms standard VQC approaches across a diverse set of tasks, including quantum dot classification, Max-Cut optimization, and molecular ground-state estimation. Of particular significance is its performance in quantum chemistry simulations, where TensorHyper-VQC achieves near-chemical-accuracy solutions for the LiH molecule with over $2.7\times$ fewer parameters than a conventional VQE. Moreover, the framework exhibits remarkable robustness under NISQ-relevant noise conditions and on real quantum hardware, maintaining low energy errors without requiring explicit error mitigation techniques. These findings establish TensorHyper-VQC as a practically deployable solution that remains stable and accurate even under realistic device noise and decoherence. 

Importantly, TensorHyper-VQC is hardware-agnostic and requires no circuit-level modifications or additional mitigation overhead, making it readily compatible with current quantum platforms. Its demonstrated robustness, coupled with strong theoretical guarantees, positions the framework as a promising foundation for scalable, noise-resilient quantum learning on near-term devices.

Beyond its immediate algorithmic contributions, TensorHyper-VQC also holds broader implications for the future of quantum computing. By enabling more efficient, scalable variational models, this approach can accelerate advances across a wide range of scientific and technological domains, from drug discovery and materials design to energy optimization and logistics. Furthermore, integrating classical tensor network techniques with quantum computational paradigms paves the way for next-generation hybrid models that harness the complementary strengths of both quantum and classical computing. Such models have the potential to transform not only quantum machine learning but also the broader landscape of scientific computing in the NISQ era and beyond.

\section{Methods}
This section provides methodological support for the theoretical analysis presented in the main text, including the NTK analysis, the concepts of TTN, QAOA, and VQE, and detailed proofs of our theoretical results. 

\subsection{An Introduction to Neural Tangent Kernel Theory} 
The Neural Tangent Kernel (NTK)~\cite{bietti2019inductive, jacot2018neural, liu2022representation} is a powerful theoretical framework that explains the efficient learnability of deep neural networks despite their inherent complexity. Fundamentally, the NTK theory examines how marginal changes in a network's parameters affect its predictions. A network is deemed "well-conditioned," leading to smoother training and faster convergence, when minor parameter adjustments reliably yield stable, predictable output shifts.

The NTK quantifies this stability by constructing a kernel that serves as a similarity measure and characterizes the network's behavior throughout training. A network endowed with a "well-conditioned" NTK, whose smallest eigenvalue is sufficiently large, will exhibit stable gradients during training, thereby enabling efficient optimization through gradient-based algorithms. Conversely, networks plagued by poorly conditioned NTKs can exhibit unstable convergence, a phenomenon often referred to as optimization difficulties.

Our theoretical framework for TensorHyper-VQC directly leverages NTK theory to analyze its trainability. The VQC and the TT network determine the NTK matrix for TensorHyper-VQC. A key theoretical finding is that the smallest eigenvalue of the NTK of TensorHyper-VQC is strictly greater than that of a conventional VQC. This indicates a more optimized landscape for TensorHyper-VQC, leading to faster, more stable convergence by mitigating the vanishing-gradient problem. Thus, the structured parameterization afforded by the TT network directly improves trainability, as supported by our NTK analysis.

\subsection{Tensor-Train Network and TTN+VQC}
The TT network~\cite{oseledets2011tensor}, also known as the matrix product state~\cite{cirac2021matrix}, is one of the most widely used tensor network architectures in both machine learning and quantum many-body physics. It provides an efficient way to represent and manipulate high-dimensional data by decomposing a large tensor into a sequence of lower-order tensors (called TT cores) arranged in a one-dimensional chain.

Formally, the TT decomposition is a structured factorization technique that represents a high-dimensional tensor as a sequence of low-rank tensor cores. Concretely, a tensor $\mathcal{G} \in \mathbb{R}^{d_1 \times d_2 \times \cdot\cdot\cdot \times d_{K}}$ is factorized as:
\begin{equation}
\mathcal{G}(i_1, i_2, ..., i_K) = \mathcal{G}_{1}(i_1) \mathcal{G}_{2}(i_2)\cdot\cdot\cdot \mathcal{G}_{K}(i_{K}),
\end{equation}
where each TT-core $\mathcal{G}_{k}(i_{k}) \in \mathbb{R}^{r_{k-1} \times r_{k}}$ is of relatively low dimension, with $(r_{0}, r_{1}, ..., r_{K})$ denoting the TT-ranks and boundary conditions $r_0 = r_K = 1$. 

This representation reduces the number of free parameters from exponential in $K$ to linear, while still capturing rich multi-way correlations among indices. This factorization dramatically reduces the parameter complexity from exponential to linear scaling with respect to the tensor order, while still capturing essential correlations among dimensions. As a result, TT networks have become a foundational tool for tasks such as dimensionality reduction, feature compression, and structured parameterization in machine learning models.

When integrated with VQC, the TT network forms a hybrid quantum-classical architecture, traditionally referred to as TTN+VQC~\cite{qi2023theoretical}. In this framework, the TT network typically serves as a classical preprocessing or feature-compression module that reduces the input dimensionality before quantum encoding. The VQC then processes the compressed features, leveraging quantum superposition and entanglement to enhance the model’s expressivity and capture nonlinear correlations that are difficult to represent classically. This complementary interplay positions TTN+VQC as a promising paradigm for tackling complex and high-dimensional quantum machine learning tasks.

\subsection{Quantum Approximate Optimization Algorithm}
The Quantum Approximate Optimization Algorithm (QAOA)~\cite{zhou2020quantum, zhu2022adaptive} is a hybrid quantum-classical algorithm designed to find approximate solutions to combinatorial optimization problems. It iteratively optimizes the parameters of a quantum circuit to maximize an objective function defined by the problem. One common application of QAOA is solving the Max-Cut problem. The Max-Cut problem seeks to partition the vertices of a graph into two sets, thereby maximizing the number of edges between them. 

In the QAOA framework, this problem is encoded as an eigenvalue problem of the quantum Hamiltonian. The algorithm then variationally optimizes the parameters of a quantum circuit to find a quantum state that, upon measurement, yields a classical bit string that approximates the optimal cut. For instance, our work demonstrates the application of QAOA to 20-qubit Erdős-Rényi graphs for the Max-Cut problem. 

\subsection{Variational Quantum Eigensolver}
The Variational Quantum Eigensolver (VQE)~\cite{kandala2017hardware, zhang2022variational} is a prominent hybrid quantum-classical algorithm designed to find the ground state energy of a given Hamiltonian. It works by variably optimizing the parameters of a quantum circuit to prepare a quantum state. The expectation value of the Hamiltonian is then measured for this state on a quantum computer and fed back to a classical optimizer. The classical optimizer then updates the quantum circuit's parameters to minimize energy, iteratively searching for the quantum state corresponding to the lowest possible energy, i.e., the ground state. VQE is a promising algorithm for NISQ devices due to its relatively shallow circuit depth requirements.

\subsection{TT Network Input and Feature Processing}

In the TensorHyper-VQC framework, the classical TT network is a hypernetwork that maps problem-specific classical features to the full quantum circuit parameters required for the target VQC task. The TT network takes as input a vectorized representation of the charge stability diagram to classify quantum dots. Specifically, each 50 × 50 image is flattened into a 2,500-dimensional real-valued vector, which is then projected or reshaped as needed to match the input dimension structure of the TT-cores. For combinatorial optimization tasks, such as Max-Cut, the TT network input consists of concise task-level descriptors, including the degree histogram and other graph statistics. 

These classical features are likewise arranged to match the input dimension requirements of each TT-core. The TT network processes this input via a sequence of low-rank tensor contractions, effectively learning a nonlinear, structured mapping from the classical feature space to the quantum circuit parameter space. The resulting output vector defines all variational parameters (e.g., rotation angles) for the quantum gates in the VQC, thereby conditioning the quantum circuit on the classical data instance or problem structure. This mechanism enables TensorHyper-VQC to dynamically adapt the quantum ansatz parameters to each input, thereby enhancing task specificity and generalization.

\subsection{Ansatz Selection of Quantum Circuits}

A distinctive feature of the TensorHyper-VQC framework is that, during training, the quantum circuit ansatz remains fixed and serves purely as a forward-pass evaluator. At the same time, all trainable parameters are generated classically by the TT hypernetwork. This design choice significantly simplifies the optimization landscape, mitigates barren plateaus, and enables robust, hardware-agnostic training on NISQ devices. Moreover, by decoupling parameter learning from circuit optimization, TensorHyper-VQC achieves stable and efficient convergence without requiring repeated quantum recompilation or topology searches.

However, this fixed-ansatz paradigm introduces an inherent trade-off between trainability and adaptability. While the TT hypernetwork can flexibly adapt parameter values within a given circuit structure, the expressive power and generalization capability of the entire model are ultimately bounded by the representational capacity of the pre-selected ansatz. In particular, when facing tasks that fundamentally differ from the training distribution, such as those requiring distinct entanglement patterns, deeper circuit depths, or alternative gate topologies, a fixed ansatz may become suboptimal, resulting in degraded model performance.

In such cases, deploying a new TensorHyper-VQC instance with a redesigned circuit architecture may be necessary to maintain high expressivity and performance. This highlights a natural direction for future work: integrating adaptive or automated ansatz selection into the TensorHyper-VQC framework. Such an extension could combine the advantages of TT-guided parameter generation with structure-level adaptability, enabling the system to dynamically reconfigure its circuit topology in response to changes in task complexity, data distribution, or hardware constraints.

\subsection{IBM Quantum Hardware: Heron r2 Processor}

All real-device experiments in this work were conducted on IBM’s Heron r2 quantum processor, a state-of-the-art superconducting quantum processing unit featuring $156$ physical transmon qubits. The Heron r$2$ belongs to IBM’s next-generation “Heron” family, which focuses on improved connectivity, reduced cross-talk, and enhanced two-qubit gate fidelities compared with previous Eagle-class devices. The processor employs a tunable-coupler architecture that optimizes qubit-to-qubit interactions and enables high-fidelity entangling operations, achieving typical two-qubit gate errors below $10^{-3}$ and CLOPS exceeding $2.0 \times 10^{5}$. With its mid-scale qubit count and low-noise properties, the Heron r2 is well-suited for benchmarking variational quantum algorithms on NISQ hardware. In our experiments, TensorHyper-VQC was deployed directly on the Heron r2 backend to validate its trainability, expressivity, and noise-resilience under realistic hardware conditions. In particular, each experimental configuration was executed with $8192$ shots per circuit, and all measurement data were aggregated through IBM Qiskit's mitigation module to reduce residual readout bias.

\subsection{The proof of Upper Bound on Approximation Capability}

Starting from the risk decomposition of approximation error $\epsilon_{\rm app}$, by $K_{\ell}$-Lipschitz continuity of the loss function $\mathcal{R}(\cdot)$, we have: 
\begin{equation}
\epsilon_{\rm app} = \mathcal{R}(h^{*}) - \mathcal{R}(f_{\boldsymbol{\theta}^{*}}) \le K_{\ell} \lVert h^{*}(\textbf{x}) - f_{\boldsymbol{\theta}^{*}}(\textbf{x}) \rVert_{1}. 
\end{equation}

After inserting intermeidate optimal VQC $f_{\textbf{w}^{*}}(\textbf{x})$, by the triangle inequality: 
\begin{equation}
\epsilon_{\rm app} \le K_{\ell}\left( \lVert h^{*}(\textbf{x}) - f_{\textbf{w}^{*}}(\textbf{x}) \rVert_{1} + \lVert f_{\textbf{w}^{*}}(\textbf{x}) - f_{\boldsymbol{\theta}^{*}}(\textbf{x}) \rVert_{1} \right).
\end{equation}

As for the first term $\lVert h^{*}(\textbf{x}) - f_{\textbf{w}^{*}}(\textbf{x}) \rVert_{1}$, it is related to the expresssivity limitation of VQC. Based on our derived theoretical results in \cite{qi2023theoretical}, for an $L$-depth and $U$-qubit VQC, we have:  
\begin{equation}
\lVert h^{*}(\textbf{x}) - f_{\textbf{w}^{*}}(\textbf{x}) \rVert_{1} \le \mathcal{O}\left(	e^{-\alpha L} \right) + \mathcal{O}\left( \frac{1}{2^{\beta U}} \right). 
\end{equation}

The second term $\lVert f_{\textbf{w}^{*}}(\textbf{x}) - f_{\boldsymbol{\theta}^{*}}(\textbf{x}) \rVert_{1}$ measures the TT parameterization error. Since VQC circuits are $K_{\textbf{w}}$-Lipschitz in their parameters (an assumption standard in VQC theory), it leads to 
\begin{equation}
\lVert f_{\textbf{w}^{*}}(\textbf{x}) - f_{\boldsymbol{\theta}^{*}}(\textbf{x}) \rVert_{1} \le K_{\textbf{w}} \lVert	 \textbf{w}^{*} - \hat{\textbf{w}} \rVert_{1}. 
\end{equation}

Moreover, for a target weight $\textbf{w}^{*}$ associated with its mode-k unfolding $\mathcal{M}_k(\textbf{w}^{*})$, the TT approximation with TT-ranks $\{r_{1}, r_2, ..., r_K\}$ (where $r_0 = r_K = 0$) satisifies the classical bound as: 
\begin{equation}
\epsilon_{\rm tt} = \lVert \textbf{w}^{*} - \hat{\textbf{w}} \rVert_{1} \le \left(\sum\limits_{k=1}^{K-1} \sum\limits_{j > r_{k}} \sigma_{j}^{2}\left( \mathcal{M}_{k}(\textbf{w}^{*}) \right)\right)^{\frac{1}{2}}, 
\end{equation}
where $\sigma_{j}\left( \mathcal{M}_{k}(\textbf{w}^{*}) \right)$ are singular values of $\mathcal{M}_{k}(\textbf{w}^{*})$. The TT approximation error $\epsilon_{\rm tt} = \epsilon_{\rm tt}(r_1, r_2, ..., r_{K-1})$ is a monotonically decreasing function of the TT-ranks, vanishing when each $r_k$ equals the full unfolding rank. 

By combining the above two terms, the approximation error bound becomes: 
\begin{equation}
\epsilon_{\rm app} \le \mathcal{O}\left(K_{\ell} e^{-\alpha L}\right) + \mathcal{O}\left( \frac{K_{\ell}}{2^{\beta U}} \right) + \mathcal{O}\left( K_{\ell} \epsilon_{\rm tt}(r_1, r_2, ..., r_{K-1})	\right). 
\end{equation}

Furthermore, let $r = \max_{1\le K-1} r_k$ be the maximum TT-rank. By the decay assumption $\sigma_{j}(\mathcal{M}_{k}(\textbf{w}^{*})) \le C_{k} j^{-p}$ and the integral test for $p > 1$, we have:
\begin{equation}
\sum\limits_{j > r_{k}} \sigma_{j}(\mathcal{M}_{k}(\textbf{w}^{*})) \le C_k \sum\limits_{j > r_k}j^{-p} \le C_k \int_{r_k}^{\infty} x^{-p} dx = \frac{C_k}{p-1}r_{k}^{-(p-1)}. 
\end{equation}

By using the Cauchy-Schwarz inequality, we obtain:
\begin{equation}
\left(\sum\limits_{k=1}^{K-1} \sum\limits_{j > r_{k}} \sigma_{j}^{2}\left( \mathcal{M}_{k}(\textbf{w}^{*}) \right)\right)^{\frac{1}{2}} \le \sum\limits_{k=1}^{K-1} \sum\limits_{j > r_{k}} \sigma_{j}\left( \mathcal{M}_{k}(\textbf{w}^{*}) \right).
\end{equation}

Then, we derive that: 
\begin{equation}
\lVert \textbf{w}^{*} - \hat{\textbf{w}} \rVert_{1} \le \sum\limits_{k=1}^{K-1} \sum\limits_{j > r_{k}} \sigma_{j}(\mathcal{M}_{k}(\textbf{w}^{*})) \le \frac{1}{p-1}\left(\sum\limits_{k=1}^{K-1}C_{k} \right) \max\limits_{k} r_{k}^{-(p-1)} = \sum\limits_{k=1}^{K}\frac{C_{k}}{p-1} r^{-(p-1)}. 
\end{equation}

By defining the TT parameterization error as $\epsilon_{\rm tt}(r):= \lVert \textbf{w}^{*} - \hat{\textbf{w}} \rVert_{1}$, we finally obtain: 
\begin{equation}
\epsilon_{\rm tt}(r) = \mathcal{O}(r^{-(p-1)}), 
\end{equation}
which corresponds to Corollary~\ref{ref:cor1}.

\subsection{Proof of Upper Bound on Generalization Performance}

The upper bound on the generalization gap arises from empirical process theory, where the generalization error of a Lipschitz loss function is upper bounded by the empirical Rademacher complexity $\mathcal{C}_{N}(\mathcal{F}_{\boldsymbol{\theta}})$ with a TensoMeta-VQC class $\mathcal{F}_{\boldsymbol{\theta}}$. 

Moreover, the training of quantum circuits introduces quantum measurement noise. In TensorHyper-VQC, the noise only appears in the loss measurement, which presents the variance term $\text{Var}[\frac{\partial \mathcal{R}(f_{\boldsymbol{\theta}})}{\partial \boldsymbol{\theta}}]$. Thus, the generalization gap can be represented as: 
\begin{equation}
\operatorname{E}\left[	\sup\limits_{f_{\boldsymbol{\theta}} \in \mathcal{F}_{\boldsymbol{\theta}}} \vert \mathcal{R}(f_{\boldsymbol{\theta}}) - \hat{\mathcal{R}}(f_{\boldsymbol{\theta}}) \vert \right] \le 2 \mathcal{C}_{N}(\mathcal{F}_{\boldsymbol{\theta}}) + \left(\text{Var}\left[ \frac{\partial \mathcal{R}(f_{\boldsymbol{\theta}})}{\partial \boldsymbol{\theta}} \right]\right)^{\frac{1}{2}}.
\end{equation}

For the functional operators in the parameter space with the NTK $\mathcal{T}_{\rm tv}$, the empirical Rademacher complexity scales as: 
\begin{equation}
\mathcal{C}_{N}(\mathcal{F}_{\boldsymbol{\theta}}) =\tilde{O}\left( \sqrt{\frac{\text{Tr}(\mathcal{T}_{\rm tv})}{N^{2}}} \right). 
\end{equation}

On the other hand, suppose the quantum measurement noise is represented by a random vector $\boldsymbol{\tau} \in \mathbb{R}^{3U}$ with mean $\textbf{b} = \operatorname{E}[\boldsymbol{\tau}]$ and covariance matrix $\boldsymbol{\Sigma} = \text{Cov}(\boldsymbol{\tau})$. Given a $K_{\ell}$-Lipschitz loss function and a training set $\{(\textbf{x}_{n}, \textbf{y}_{n})\}_{n=1}^{N}$, using the linearized NTK approximation, we have:
\begin{equation}
\left(\text{Var}\left[ \frac{\partial \mathcal{R}(f_{\boldsymbol{\theta}})}{\partial \boldsymbol{\theta}} \right]\right)^{\frac{1}{2}} \le \mathcal{O}\left( \sqrt{\frac{\lambda_{\rm max}(\boldsymbol{\Sigma}) K_{\ell}^{2}}{N}} \right) + \mathcal{O}\left(\frac{\lVert \textbf{b} \rVert_{2} K_{\ell}}{\sqrt{N}} \right), 
\end{equation}
where $\hat{\mathcal{R}}(\cdot)$ denotes the empirical risk, and $\lambda_{\rm max}(\boldsymbol{\Sigma})$ is the maximum eigvenvalue of $\boldsymbol{\Sigma}$. 

Putting all the pieces together, we obtain: 
\begin{equation}
\operatorname{E}\left[	\sup\limits_{\boldsymbol{\theta}} \left\vert \mathcal{R}(f_{\boldsymbol{\theta}}) - \hat{\mathcal{R}}(f_{\boldsymbol{\theta}}) \right\vert \right] \le \tilde{O}\left(\sqrt{\frac{\text{Tr}(\mathcal{T}_{\rm tv})}{N^{2}}}\right) + \mathcal{O}\left( \sqrt{\frac{\lambda_{\rm max}(\boldsymbol{\Sigma}) K_{\ell}^{2}}{N}} \right) + \mathcal{O}\left(\frac{\lVert \textbf{b} \rVert_{2} K_{\ell}}{\sqrt{N}}	\right), 
\end{equation}
which matches our theoretical result in Generalization Performance. 

Furthermore, for a parameterized model $f_{\boldsymbol{\theta}}(\textbf{x})$, the NTK $\mathcal{T}_{\rm tv}$ is: 
\begin{equation}
\mathcal{T}_{\rm mv}\left(\textbf{x}, \textbf{x}' \right) = \langle \nabla_{\boldsymbol{\theta}}f_{\boldsymbol{\theta}}(\textbf{x}), \nabla_{\boldsymbol{\theta}}f_{\boldsymbol{\theta}}(\textbf{x}') \rangle. 
\end{equation}

The trace of NTK $\mathcal{T}_{\rm tv}$ over a dataset $\{\textbf{x}_{i}\}_{n=1}^{N}$ is: 
\begin{equation}
\text{Tr}(\mathcal{T}_{\rm tv}) = \sum\limits_{n=1}^{N} \mathcal{T}_{\rm mv}(\textbf{x}_{n}, \textbf{x}_{n}) = \sum\limits_{n=1}^{N} \lVert \nabla_{\boldsymbol{\theta}} f_{\boldsymbol{\theta}}(\textbf{x}_{n}') \rVert_{2}^{2}.
\end{equation}

For TensorHyper-VQC, the model output is: 
\begin{equation}
f_{\boldsymbol{\theta}}(\textbf{x}) = \operatorname{VQC}\left( \hat{\textbf{w}}(\textbf{x}; \boldsymbol{\theta}) \right), 
\end{equation}
where $\hat{\textbf{w}}(\textbf{x}; \boldsymbol{\theta})$ is generated by a TT network from TT-cores $\boldsymbol{\theta}$. 

By using the chain rule: 
\begin{equation}
\nabla_{\boldsymbol{\theta}}f_{\boldsymbol{\theta}}(\textbf{x}) = \frac{\partial f_{\boldsymbol{\theta}}}{\partial \hat{\textbf{w}}} \frac{\partial \hat{\textbf{w}}}{\partial \boldsymbol{\theta}},
\end{equation}
where $\frac{\partial f}{\partial \hat{\textbf{w}}}$ is determined bhy the quantum circuit, and $\frac{\partial \hat{\textbf{w}}}{\partial \boldsymbol{\theta}}$ is highly structured because: (1) Each output parameter depends on only a small subset of TT-cores due to the TT's multilinear structure; (2) This introduces strong correlations among gradient across different parameters. 

In a full parameterization, $\nabla_{\boldsymbol{\theta}}f_{\boldsymbol{\theta}}(\textbf{x})$ can be ``dense" in the parameter space, and the squared norm $\lVert \nabla_{\boldsymbol{\theta}}f_{\boldsymbol{\theta}}(\textbf{x}) \rVert_{2}^{2}$ can scale with the effective dimension of the function class which is at most $\prod_{k=1}^{K}r_{k}$. Formally, 
\begin{equation}
\lVert \nabla_{\boldsymbol{\theta}}f_{\boldsymbol{\theta}}(\textbf{x}) \rVert_{2}^{2} \le C_r \cdot \prod\limits_{k=1}^{K} r_{k}, 
\end{equation}
for some constant $C_r$ (depending on the input scaling, circuit, and TT initialization). 

Therefore, we find that the generalization gap is controlled polynomially in TT-rank, which is:
\begin{equation}
\text{Tr}(\mathcal{T}_{\rm mv}) = \sum\limits_{n=1}^{N} \lVert \nabla_{\boldsymbol{\theta}} f_{\boldsymbol{\theta}}(\textbf{x}_{n}') \rVert_{2}^{2}\le NC_r \cdot  \prod\limits_{k=1}^{K} r_{k}. 
\end{equation}

\subsection{Proof of Robustness to Quantum Noise}

The key observation is that in a standard VQC, every parameter (i.e., gate angle) directly receives noise from quantum hardware measurements. For a circuit with $3UL$ parameters (where $U$ is the number of qubits and $L$ is the number of layers), each parameter's gradient is independently corrupted, causing the total noise to grow proportionally with system size. As a result, large-scale VQCs become highly unstable in the presence of hardware noise. By contrast, TensorHyper-VQC introduces a classical TT hypernetwork that maps TT-cores $\{\mathcal{G}_k\}_{k=1}^{K}$ to gate parameters $\hat{\textbf{w}}$. This mapping has two fundamental consequences: (i) it spreads gradient sensitivity across many TT-core elements, so no single parameter is overly sensitive to noise, and (ii) it premultiplies the noise by the Jacobian $\textbf{J}_k^\top$, which naturally shrinks with the number of qubits and layers. 

These properties imply that the noise entering the gradient computation is attenuated by a factor that decreases as $1/\sqrt{UL}$ or $1/(UL)$, effectively acting as an intrinsic noise filter. 

We first consider noise propagation in gradient computation. Let $\hat{\textbf{w}}\in\mathbb{R}^{3UL}$ denote all quantum gate parameters and $\{\mathcal{G}_k\}_{k=1}^{K}$ the TT cores. The Jacobian $\textbf{J}_k$ is defined as: 
\begin{equation}
\textbf{J}_k = \frac{\partial \hat{\textbf{w}}}{\partial\, \mathcal{G}_k} \in \mathbb{R}^{(3UL)\times d_k}.
\end{equation} 

Moreover, measurement noise in angle space can be written as: 
\begin{equation}
\frac{\partial \hat{\mathcal{R}}}{\partial \hat{\textbf{w}}} = \frac{\partial \mathcal{R}}{\partial \hat{\textbf{w}}} + \boldsymbol{\tau},
\end{equation} 
where $\boldsymbol{\tau}$ denotes measurement noise with mean $\textbf{b}$ and covariance $\boldsymbol{\Sigma}$. Using the chain rule, the gradient with respect to the TT core is: 
\begin{equation}
  \frac{\partial \hat{\mathcal{R}}}{\partial \mathcal{G}_k} = \textbf{J}_k^\top \frac{\partial \hat{\mathcal{R}}}{\partial \hat{\textbf{w}}} = \underbrace{\textbf{J}_k^\top \frac{\partial \mathcal{R}}{\partial \hat{\textbf{w}}}}_{\textbf{g}_k} + \textbf{J}_k^\top \boldsymbol{\tau}.
\end{equation}
This decomposition shows that noise always appears multiplied by $\textbf{J}_{k}^{\top}$. 

The first key property is that the TT Jacobian $\textbf{J}_{k}$ has its sensitivity distributed evenly across all $3UL$ outputs. Under standard TT initialization and mild assumptions, each partial derivative $\frac{\partial \hat{w}_u}{\partial g_{k, j}}$ is zero-mean with variance $\sigma_{k}^{2}$ independent of $u$ and $j$. This gives:
\begin{equation}
\lVert	 \textbf{J}_k \rVert_{2}^{2} \lesssim \frac{c}{3UL},
\end{equation} 
where $c$ is a constant and does not grow with $U$ or $L$. 

The result shows that the TT network spreads parameter influence so broadly that no single parameter contributes strongly to the gradient. This “energy dispersion” causes the Jacobian norm to shrink as the model grows. 

Suppose the noise $\boldsymbol{\tau}$ is independent, zero-mean, and has variance $\sigma^{2}$. Then, we achieve: 
\begin{equation}
\text{Var}\left[ \frac{\partial \hat{\mathcal{R}}}{\partial \mathcal{G}_{k}} \right] = \textbf{J}^{\top}_{k} (\sigma^{2} I) \textbf{J}_{k} \lesssim \sigma^{2} \lVert \textbf{J}_{k} \rVert_{2}^{2} \lesssim \frac{c\sigma^{2}}{3UL}. 
\end{equation}

The result means that even when noise components are correlated, the Jacobian still damps them by the same $\frac{1}{UL}$ scaling. Only the constant $\lambda_{\rm max}$ (the largest noise eigenvalue) matters. 

If the noise covariance $\boldsymbol{\Sigma}$ is arbitrary (non-diagonal, possibly correlated): 
\begin{equation}
\text{Var}\left[ \frac{\partial \hat{\mathcal{R}}}{\partial \mathcal{G}_{k}} \right] = \textbf{J}_k \boldsymbol{\Sigma} \textbf{J}_k \lesssim \lambda_{\rm max}(\boldsymbol{\Sigma}) \lVert \textbf{J}_{k} \rVert_{2}^{2} \lesssim \frac{c\lambda_{\rm max}(\boldsymbol{\Sigma})}{3UL}. 
\end{equation}

Even when noise components are correlated, the Jacobian still damps them by the same $\frac{1}{UL}$ scaling. Only the constant $\lambda_{\rm max}$ (the largest noise eigenvalue) matters. 

If the covariance is diagonally dominant (e.g., readout noise where each measurement’s total variance is bounded by $B$): 
\begin{equation}
\text{Var}\left[ \frac{\partial \hat{\mathcal{R}}}{\partial \mathcal{G}_{k}} \right]  \lesssim \frac{c B}{3UL},
\end{equation}

which suggests that even with structured, non-i.i.d. noise, the variance stills shrinks inversely with $UL$. This shows that robustness is not an artifact of simple noise assumptions associated with a structural property. 

If the noise has a non-zero mean and arbitrary covariance, then the gradient estimate becomes: 
\begin{equation}
\hat{\textbf{g}}_{k} = \textbf{g}_k + \textbf{J}_{k}^{\top} \boldsymbol{\tau}. 
\end{equation}

The bias and variance are: 
\begin{equation}
\operatorname{E}[\hat{\textbf{g}}_{k}] - \textbf{g}_{k} = \textbf{J}_{k}^{\top} \textbf{b}, \hspace{2.5mm} \text{Var}(\hat{\textbf{g}}_{k}) \lesssim \frac{c\lambda_{\rm max}(\boldsymbol{\Sigma})}{3UL}. 
\end{equation}

The result demonstrates that both bias and variance shrink with system size: the bias decreases as $\frac{1}{\sqrt{UL}}$, and the variance decreases as $\frac{1}{UL}$. Even systematic measurement errors become less harmful as the circuit grows. 

In comparison with the standard VQC, if we parameterize the circuit directly by $\textbf{w}$ (no TT mapping), we attain: 
\begin{equation}
\hat{\textbf{g}}_{\textbf{w}} = \frac{\partial \hat{\mathcal{R}}}{\partial \textbf{w}} = \frac{\partial \mathcal{R}}{\partial \textbf{w}} + \boldsymbol{\tau}. 
\end{equation}

Then, we have:
\begin{equation}
\text{Var}(\hat{\textbf{g}}_{\textbf{w}}) = \boldsymbol{\Sigma}, \hspace{3mm} \operatorname{E}[\hat{\textbf{g}}_{\textbf{w}}] - \textbf{g}_{\textbf{w}} = \textbf{b},
\end{equation}
which means that the noise variance and bias are independent of $U$ or $L$. As circuits scale, noise accumulates rather than averaging out. By contrast, in TensorHyper-VQC, the Jacobian naturally filters the noise. This gives the system an inherent robustness that improves with scale.

\section{Acknowledgements}
This work is partly funded by the Hong Kong Research Impact Fund (R6010-23).

\section{Data Availability Statement}
The dataset used in our experiments on quantum dot classification is available at https://gitlab.com/QMAI/mlqe$\_$2023$\_$edx.

\section{Code Availability Statement}
Our code for TensorHyper-VQC and other VQC models is available on the website: https://github.com/jqi41/TensorHyper.  

\section{Competing Interests}
The authors declare no Competing Financial or Non-Financial Interests.

\section{Author Contributions}
Jun Qi and Chao-Han Yang conceived the project. Jun Qi and Min-Hsiu Hsieh completed the theoretical analysis. Jun Qi, Chao-Han, and Pin-Yu Chen designed the experimental work. Min-Hsiu Hsieh and Pin-Yu Chen provided high-level advice on the paperwork pipeline, and Jun Qi wrote the manuscript. 

\section{References}

\bibliographystyle{IEEEbib}
\bibliography{sn-bibliography}

\end{document}